\documentclass[final,3p,times,twocolumn,numbers]{elsarticle}

\usepackage{amssymb}
\usepackage{amsthm}
\usepackage{amsmath}
\usepackage{mathtools}
\usepackage{booktabs}
\usepackage{mhchem}
\usepackage{xcolor}
\usepackage{csvsimple,booktabs}
\usepackage{graphicx}
\usepackage{lscape}
\usepackage{algorithm}
\usepackage{algorithmicx}
\usepackage{algpseudocode}
\usepackage{adjustbox}
\usepackage[counterclockwise, figuresleft]{rotating}
\usepackage{pdflscape}

\usepackage{subcaption}
\usepackage{mwe}

\journal{Sustainable Computing: Informatics and Systems}

\begin{document}

\begin{frontmatter}


 \title{The impact of online machine-learning methods on long-term investment decisions and generator utilization in electricity markets}
 \author{Alexander J. M. Kell}
 \ead{a.kell2@newcastle.ac.uk}



\author{A. Stephen McGough, Matthew Forshaw}

\address{School of Computing, Newcastle University, Newcastle upon Tyne, United Kingdom}

\begin{abstract}


Electricity supply must be matched with demand at all times. This helps reduce the chances of issues such as load frequency control and the chances of electricity blackouts. To gain a better understanding of the load that is likely to be required over the next 24 hours, estimations under uncertainty are needed. This is especially difficult in a decentralized electricity market with many micro-producers which are not under central control. 


In this paper, we investigate the impact of eleven offline learning and five online learning algorithms to predict the electricity demand profile over the next 24 hours. We achieve this through integration within the long-term agent-based model, ElecSim. Through the prediction of electricity demand profile over the next 24 hours, we can simulate the predictions made for a day-ahead market. Once we have made these predictions, we sample from the residual distributions and perturb the electricity market demand using the simulation, ElecSim. This enables us to understand the impact of errors on the long-term dynamics of a decentralized electricity market.


We show we can reduce the mean absolute error by 30\% using an online algorithm when compared to the best offline algorithm, whilst reducing the required tendered national grid reserve required. This reduction in national grid reserves leads to savings in costs and emissions. We also show that large errors in prediction accuracy have a disproportionate error on investments made over a 17-year time frame, as well as electricity mix.

\end{abstract}

%
%

\begin{keyword}
Online learning \sep Machine learning \sep Market investment \sep Climate Change \sep Machine Learning


\end{keyword}

\end{frontmatter}


\section{Introduction}
\label{sec:intro}

The integration of higher proportions of intermittent renewable energy sources (IRES) in the electricity grid will mean that the forecasting of electricity supply and demand will become increasingly challenging \footnote{Matching of supply and demand will be referenced as demand from this point onwards for brevity.}. Examples of IRES are solar panels and wind turbines. These fluctuate in terms of power output based on localized wind speed and solar irradiance. As supply must meet demand at all times and the fact that IRES are less predictable than dispatchable energy sources such as coal and combined-cycle gas turbines (CCGTs), extra attention must be made in predicting future demand if we wish to keep, or reduce, the current frequency of blackouts \cite{Lu1993}. A dispatchable source is one that can be turned on and off by human control and is able to adjust output just in time at a moment convenient for the grid.

Typically, peaker plants, such as reciprocal gas engines, are used to fill fluctuations in demand that had not been previously planned for. Peaker plants meet the peaks in demand when other cheaper options are at full capacity. Current peaker plants are expensive to run and have higher greenhouse gas emissions than their non-peaker counterparts. Whilst peaker plants are also dispatchable plants, not all dispatchable plants are peaker plants. For example coal, which is a dispatchable plant, is run as a base load plant, due to its inability to deal with fluctuating conditions.

To reduce reliance on peaker plants, it is helpful to know how much electricity demand there will be in the future so that more efficient plants can be used to meet this expected demand. This is so that these more efficient plants can be brought up to speed at a time suitable to match the demand. Forecasting a day into the future is especially useful in decentralized electricity markets which have day-ahead markets. Decentralized electricity markets are ones where electricity is provided by multiple generation companies, as opposed to a centralized source, such as a government. To aid in this prediction, machine learning and statistical techniques have been used to accurately predict demand based on several different factors and data sources \cite{Kell2018a}, such as weather \cite{Hong2014}, day of the week \cite{Al-Musaylh2018} and holidays \cite{Vrablecova2017}. 



Various studies have looked at predicting electricity demand at various horizons, such as short-term \cite{Huang2003} and long-term studies \cite{Andersen2013}. However, the impact of poor demand predictions on the long-term electricity mix has been studied to a lesser degree.

In this paper, we compare several machine learning and statistical techniques to predict the energy demand for each hour over the next 24-hour horizon. We chose to predict over the next 24 hours to simulate a day-ahead market, which is often seen in decentralized electricity markets. However, our approach could be utilized for differing time horizons. In addition to this, we use our long-term agent-based model, ElecSim \cite{Kell, Kell2020}, to simulate the impact of different forecasting methods on long-term investments, power plant usage and carbon emissions between 2018 and 2035 in the United Kingdom. Our approach, however, is generalizable to any country through parametrization of the ElecSim model.

Within energy modelling different methodologies are undertaken to understand how an energy system may develop. In this paper we used an agent-based model (ABM). An other approach is the use of optimization based models. Optimization based models develop scenarios into the future by finding a cost-optimal solution. These models rely on perfect information of the future, assume that there is a central planner within an energy market and are solved in a normative manner. That is, how a problem should be solved under a specific scenario. Agent-based models on the other hand can assume imperfect information and allows a scenario to develop. We believe that ABMs closely match real-life in decentralised electricity markets. 



As part of our work, we utilize online learning methods to improve the accuracy of our predictions. Online learning methods can learn from novel data while maintaining what was learnt from previous data. Online learning is useful for non-stationary datasets, and time-series data where recalculation of the algorithm would take a prohibitive amount of time. Offline learning methods must be retrained every time new data is added. Online approaches are constantly updated and do not require significant pauses. 

We trial different learning algorithms for different times of the year. Specifically, we train different algorithms for the different seasons. We also split weekdays and train both weekends and holidays together. This is due to the fact that holidays and weekends exhibit similar load profiles due to the reduction in industrial electricity use and an increase in domestic use. This enables an algorithm to become good at a specific subset of the data which share similar patterns, as opposed to having to generalize to all of the data. Examples of the algorithms used are linear regression, lasso regression, random forests, support vector regression, multilayer perceptron neural network and the box-cox transformation. 

We expect a-priori that online algorithms will outperform the offline approach. This is due to the fact that the demand time-series is non-stationary, and thus changes sufficiently over time. In terms of the algorithms, we presume that the machine learning algorithms, such as neural networks, support vector regression and random forests will outperform the statistical methods such as linear regression, lasso regression and box-cox transformation regression. This is due to the fact that machine learning has been shown to be able to learn more complex feature representations than statistical methods \cite{Singh2012}. In addition, our previous work has shown that the random forest was able to outperform neural networks and support vector regression \cite{Kell2018}. 

It should be noted, that such a-priori intuition, is no substitute for analytical evidence and can (and has) been shown to be wrong in the past, due to imperfect knowledge of the data and understanding of some of the black box algorithms, such as neural networks.


Using online and offline methods, we take the error distributions, or residuals, and fit a variety of distributions to these residuals. We choose the distribution with the lowest sum of squared estimate of errors (SSE). SSE was chosen as the metric to ensure that both positive and negative errors were treated equally, as well as ensuring that large errors were penalized more than smaller errors. We fit over 80 different distributions, which include the Johnson Bounded distribution, the uniform distribution and the gamma distribution. The distribution that best fits the respective residuals is then used and sampled from to adjust the demand in the ElecSim model. We then observe the differences in carbon emissions, and which types of power plants were both invested in and utilized, with each of the different statistical and machine learning methods. To the best of our knowledge, this is the most comprehensive evaluation of online learning techniques to the application of day-ahead load forecasting as well as assessing the impacts of the errors that these algorithms produce on the long-term electricity market dynamics.


We show that online learning has a significant impact on reducing the error for predicting electricity consumption a day ahead when compared to traditional offline learning techniques, such as multilayer perceptron artificial neural networks, linear regression, extra trees regression and support vector regression, which are algorithms used in the literature \cite{Lu1993, Ahmad2017, Chen2004}. 

We show that the forecasting algorithm has a non-negligible impact on carbon emissions and use of coal, onshore, photovoltaics (PV), reciprocal gas engines and CCGT. Specifically, the amount of coal, PV, and reciprocal gas used from 2018 to 2035 was proportional to the median absolute error, while wind is inversely proportional to the median absolute error.

Total investments in coal, offshore and photovoltaics are proportional to the median absolute error, while investments in CCGT, onshore and reciprocal gas engines are inversely proportional. 

In this work, we make the following contributions:

\begin{enumerate}
  \item The evaluation of different online and offline learning algorithms to forecast the electricity demand profile 24 hours ahead. This work extends previous work by utilizing a vast array of different online and offline techniques.
  \item Evaluation of poor predictive ability on the long-term electricity market in the UK through the perturbation of demand in the novel ElecSim simulation. There remains a gap in the literature of the long-term impact of poor electricity demand predictions on the electricity market.
\end{enumerate}


In Section \ref{sec:lit-review}, we review the literature. We introduce the dynamics of the ElecSim simulation as well as the methods used in Section \ref{sec:material}. We demonstrate the methodology undertaken in Section \ref{sec:methods}. In Section \ref{sec:results} we demonstrate our results, followed by a discussion in Section \ref{sec:discussion}. We conclude our work in Section \ref{sec:conclusion}.

\section{Literature Review}
\label{sec:lit-review}

Multiple papers have looked at demand-side forecasting \cite{Singh2012}. These include both artificial intelligence \cite{Kim2000, Tiong2008,Quilumba2014} and statistical techniques \cite{Huang2003,Nguyen2017}. In addition to this, our research models the impact of the performance of different algorithms on investments made, electricity sources dispatched and carbon emissions over a 17 year period. To model this, we use the long-term electricity market agent-based model, ElecSim \cite{Kell2020}.

\subsection{Offline learning}

Multiple electricity demand forecasting studies have been undertaken for offline learning \cite{Chen2004, Gross1987, Ghofrani}. Studies have been undertaken using both smart meter data, as well as with aggregated demand, similar to the work in this paper. Smart meters are a type of energy meter installed in each house, which monitor electricity usage at short intervals, such as every 15 or 30 minutes. All of the papers reviews in this subsection focus on offline learning, as opposed to online learning like in this work. Additionally, we look at the long-term impact of poor predictions on the electricity market.

Fard \textit{et al.} propose a new hybrid forecasting method based on the wavelet transform, autoregressive integrated moving average (ARIMA) and artificial neural network (ANN) \cite{Fard2014}. The ARIMA method is utilized to capture the linear component of the time series, with the residuals containing the non-linear components. The non-linear parts are decomposed using the discrete wavelet transform which finds the sub-frequencies. These residuals are then used to train an ANN  to predict the future residuals. The ARIMA and ANNs outputs are then summed. Their results show that this technique can improve forecasting results.

Humeau \textit{et al.} compare MLPs, SVRs and linear regression at forecasting smart meter data \cite{Humeau2013}. They aggregate different households and observe which algorithms work the best. They find that linear regression outperforms both MLP and SVR when forecasting individual households. However, after aggregating over 32 households, SVR outperforms linear regression.

Quilumba \textit{et al.} also apply machine learning techniques to individual households' electricity consumption by aggregation \cite{Fard2014}. To achieve this aggregation, they use \textit{k}-means clustering to aggregate the households to improve their forecasting ability. The authors also use a neural network based algorithm for forecasting, and show that the number of optimum clusters for forecasting is dependent on the data, with three clusters optimal for a particular dataset, and four for another.

In our previous work, we evaluate the performance of ANNs, random forests, support vector regression and long short-term memory neural networks \cite{Kell2018a}. We utilize smart meter data, and cluster by household using the k-means clustering algorithm to aggregate groups of demand. Through this clustering we are able to reduce the error, with the random forest performing the best.

\subsection{Online learning}

There have been several studies in diverse applications on the use of online machine learning to predict time-series data, however, to the best of our knowledge there are limited examples where this is applied to electricity markets. In our work, we trial a range of algorithms to our problem. Due to time constraints, we do not trial the additional techniques discussed in this literature review within our paper. 

Johansson \textit{et al}. apply online machine learning algorithms for heat demand forecasting \cite{Johansson2017}. They find that their demand predictions display robust behaviour within acceptable error margins. They find that artificial neural networks (ANNs) provide the best forecasting ability of the standard algorithms and can handle data outside of the training set. Johansson \textit{et al.}, however, do not look at the long-term effects of different algorithms, which is a central part to this work.

Baram \textit{et al.} combine an ensemble of active learners by developing an active-learning master algorithm \cite{Baram2003}. To achieve this, they propose a simple maximum entropy criterion that provides effective estimates in realistic settings. Their active-learning master algorithm is shown to, in some cases, outperform the best algorithm in the ensemble on a range of classification problems.

Schmitt \textit{et al} also extends on existing algorithms through an extension of the FLORA algorithm in \cite{Schmitt2008, Widmer1996}. The FLORA algorithm generates a rule-based algorithm, which has the ability to make binary decisions. Their FLORA-MC enhances the FLORA algorithm for multi-classification and numerical input values. They use this algorithm for an ambient computing application. Ambient computing is where computing and communication merges into everyday life. They find that their algorithm outperforms traditional offline learners. Our work focuses on electricity demand, however.

Similarly to us, Pindoriya \textit{et al}. trial several different machine learning methods such as adaptive wavelet neural network (AWNN) for predicting electricity price forecasting \cite{Pindoriya2008}. They find that AWNN has good prediction properties when compared to other forecasting techniques such as wavelet-ARIMA, multilayer perceptron (MLP) and radial basis function (RBF) neural networks. We, however, focus on electricity demand.

Goncalves Da Silva \textit{et al}. show the effect of prediction accuracy on local electricity markets \cite{GoncalvesDaSilva2014}. They compare forecasting of groups of consumers. They trial the use of the Seasonal-Naïve and Holt-Winters algorithms and look at the effect that the errors have on trading in an intra-day electricity market of consumers and prosumers. They found that with a photovoltaic penetration of 50\%, over 10\% of the total generation capacity was uncapitalized and roughly 10, 25 and 28\% of the total traded volume were unnecessary buys, demand imbalances and unnecessary sells respectively. This represents energy that the participant has no control. Uncapitalized generation capacity is where a participant could have produced energy, however, it was not sold on the market. Additionally, due to forecast errors, the participant might have sold less than it should have. Our work, however, focuses on a national electricity market, as opposed to a local market.

\section{Material}
\label{sec:material}
%
%



Examples of online learning algorithms are Passive Aggressive (PA) Regressor \cite{Gzik2014}, Linear Regression, Box-Cox Regressor \cite{Box1964}, K-Neighbors Regressor \cite{forgy65} and Multilayer perceptron regressor \cite{Hinton1989}. For our work, we trial the stated algorithms, in addition to a host of offline learning techniques. The offline techniques trialled were Lasso regression \cite{Tibshirani1996a}, Ridge regression \cite{GeladiPaul1994Mrac},  Elastic Net \cite{Geostatistics2010}, Least Angle Regression \cite{Fike1988}, Extra Trees Regressor \cite{Fike1988}, Random Forest Regressor \cite{Breiman2001}, AdaBoost Regressor \cite{Freund1997}, Gradient Boosting Regressor \cite{316} and Support vector regression \cite{Cortes1995}. We chose the boosting and random forest techniques due to previous successes of these algorithms when applied to electricity demand forecasting \cite{Kell2018}. We trialled the additional algorithms due to availability of these algorithms using the packages scikit-learn and Creme \cite{scikit-learn,creme}. 


\noindent{\textit{\textbf{Linear regression.}}} Linear regression is a linear approach to modelling the relationship between a dependent variable and one or more independent variables. Linear regressions can be used for both online and offline learning. In this work, we used them for both online and offline learning. Linear regression algorithms are often fitted using the least squares approach. The least squares approach minimizes the sum of the squares of the residuals. 

Other methods for fitting linear regressions are by minimizing a penalized version of the least squares cost function, such as in ridge and lasso regression \cite{Tibshirani1996a, GeladiPaul1994Mrac}. Ridge regression is a useful approach for mitigating the problem of multicollinearity in linear regression. Multicollinearity is where one predictor variable can be linearly predicted from the others with a high degree of accuracy. This phenomenon often occurs in algorithms with a large number of parameters. 

In ridge regression, the OLS loss function is augmented so that we not only minimize the sum of squared residuals but also penalized the size of parameter estimates, in order to shrink them towards zero:
\begin{equation}
    L_{ridge}(\hat{\beta})=\sum^n_{i=1}(y_i-x'_i\hat{\beta})^2+\lambda\sum^m_{j=1}\hat{\beta^2_j}=||y-X\hat{\beta}||^2+\lambda||\hat{\beta}||^2.
\end{equation}
Where $\lambda$ is the regularization penalty which can be chosen through cross-validation, or the value that minimizes the cross-validated sum of squared residuals, for instance. $n$ is the number of observations of the response variable, $Y$, with a linear combination of $m$ predictor variables, $X$, and we solve for $\hat{\beta}$, where $\hat{\beta}$ are the OLS parameter estimates. Where $y_i$ is the outcome and $x_i$ are the covariates, both for the $i^{th}$ case.

Lasso is a linear regression technique which performs both variable selection and regularization. It is a type of regression that uses shrinkage. Shrinkage is where data values are shrunk towards a central point, such as the mean. The lasso algorithm encourages models with fewer parameters and automates variable selection.

Under Lasso the loss is defined as:

\begin{equation}
    L_{lasso}(\hat{\beta})=\sum^n_{i=1}(y_i-x'_i\hat{\beta})^2+\lambda\sum^m_{j=1}|\hat{\beta}_j|.
\end{equation}

The only difference between lasso and ridge regression is the penalty term.

Elastic net is a regularization regression that linearly combines the penalties of the lasso and ridge methods. Specifically, Elastic Net aims to minimize the following loss function:
\begin{equation}
L_{enet}(\hat{\beta})=\frac{\sum^n_{i=1}(y_i-x'_i\hat{\beta})^2}{2n}+\lambda\left(\frac{1-\alpha}{2}\sum^m_{j=1}\hat{\beta}^2_j+\alpha\sum^m_{j=1}|\hat{\beta_j}|\right),
\end{equation}
where $\alpha$ is the mixing parameter between ridge ($\alpha=0$) and lasso ($\alpha=1$). The parameters $\lambda$ and $\alpha$ can be tuned.

Least Angle Regression (LARS) provides means of producing an estimate of which variables to include in a linear regression, as well as their coefficients.

%

\noindent{\textit{\textbf{Decision tree-based algorithms.}}} The decision tree is an algorithm which goes from observations to output using simple decision rules inferred from data features \cite{Quinlan}. To build a regression tree, recursive binary splitting is used on the training data. Recursive binary splitting is a greedy top-down algorithm used to minimize the residual sum of squares. The RSS, in the case of a partitioned feature space with $M$ partitions, is given by:

\begin{equation}
    RSS=\sum^M_{m=1}\sum_{i\in R_m}(y-\hat{y}_{R_m})^2.
\end{equation}

\noindent Where $y$ is the value to be predicted and $\hat{y}$ is the predicted value for partition $R_m$.

Beginning at the top of the tree, a split is made into two branches. This split is carried out multiple times and the split is chosen that minimizes the current RSS. To obtain the best sequence of subtrees cost complexity, pruning is used as a function of $\alpha$. $\alpha$ is a tuning parameter that balances the depth of the tree and the fit to the training data.

The AdaBoost training process selects only the features of an algorithm known to improve the predictive power of the model \cite{Freund1997}. By doing this, the dimensionality of the algorithm is reduced and can improve compute time. This can be used in conjunction with multiple different algorithms. In our paper, we utilized the decision tree based algorithm with AdaBoost.

Random Forests are an ensemble learning method for classification and regression \cite{Breiman2001}. Ensemble learning methods use multiple learning algorithms to obtain better predictive performance. They work by constructing multiple decision trees and output the predicted value that is the mode of the predictions of the trees.

To ensure that the individual decision trees within a Random Forest are not correlated, bagging is used to sample from the data. Bagging is the process of randomly sampling with replacement of the training set and fitting the trees. This has the benefit of reducing the variance of the algorithm without increasing the bias. 

Random Forests differ in one way from this bagging procedure. Namely, using a modified tree learning algorithm that selects, at each candidate split in the learning process, a random subset of the features, known as feature bagging. Feature bagging is undertaken due to the fact that some predictors with a high predictive ability may be selected many times by the individual trees, leading to a highly correlated Random Forest.

ExtraTrees adds one further step of randomization \cite{Fike1988}. ExtraTrees stands for extremely randomized trees. There are two main differences between ExtraTrees and Random Forests. Namely, each tree is trained using the whole learning sample (And not a bootstrap sample), and the top-down splitting in the tree learner is randomized. That is, instead of computing an optimal cut-point for each feature, a random cut-point is selected from a uniform distribution. The split that yields the highest score is then chosen to split the node.

\noindent{\textit{\textbf{Gradient Boosting}}}. Gradient boosting is also an ensemble algorithm \cite{316}. Gradient boosting optimizes a cost-function over function space by iteratively choosing a function that points in the negative gradient descent direction, known as a gradient descent method.

\noindent{\textit{\textbf{Support vector regression (SVR).}}} SVR is an algorithm which finds a hyperplane and decision boundary to map an input domain to an output \cite{Cortes1995}. The hyperplane is chosen by minimizing the error within a tolerance.

Suppose we have the training set: $(x_1,y_1), \ldots,(x_i,y_i),\ldots,(x_n,y_n)$, where $x_i$ is the input, and $y_i$ is the output value of $x_i$. Support Vector Regression solves an optimization problem \cite{Chen2004, Shu2006}, under given parameters $C>0$ and $\varepsilon >0$, the form of support vector regression is \cite{Drucker1997}: 

\begin{equation}
\min_{\omega,b,\xi,\xi^{*}}\frac{1}{2}\omega^T\omega+C\sum_{i=1}^{n}(\xi_i+\xi_i^*)
\end{equation}

\noindent subject to
\begin{align}
\begin{multlined}
\label{svr:constrains}
y_i-(\omega^T\phi(x_i)+b)\leq\varepsilon+\xi_i^{*},\\
(\omega^T\phi(x_i)+b)-y_i\leq\varepsilon+\xi_i,\\
\xi_i,\xi^*_i\geq0,i=1,\ldots,n
\end{multlined}
\end{align}

\noindent $x_i$ is mapped to a higher dimensional space using the function $\phi$. The $\varepsilon$-insensitive tube $(\omega^T\phi(x_i)+b)-y_i\leq\varepsilon$ is a range in which errors are permitted, where $b$ is the intercept of a linear function. $\xi_i$ and $\xi^*_i$ are slack variables which allow errors for data points which fall outside of $\varepsilon$. This enables the optimization to take into account the fact that data does not always fall within the $\varepsilon$ range \cite{Smola2004}.

The constant $C>0$ determines the trade-off between the flatness of the support vector function. $\omega$ is the algorithm fit by the SVR. The parameters which control regression quality are the cost of error $C$, the width of the tube $\varepsilon$, and the mapping function $\phi$ \cite{Shu2006,Chen2004}.

\noindent{\textit{\textbf{K-Neighbors Regressor.}}} K-Neighbors regression is a non-parametric method \cite{forgy65}. The input consists of a new data point, and the algorithm finds the \textit{k} closest training examples in the feature space. The output is the mean value of the \textit{k} nearest neighbours.

\noindent{\textit{\textbf{Multilayer perceptron.}}} A neural network can be used in offline and online cases. Here, we used them for both.

\begin{figure}
\centering
    \includegraphics[width=0.3\textwidth]{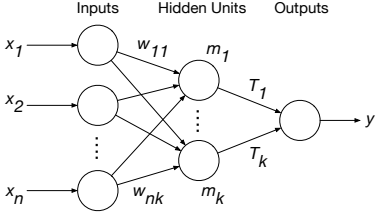}
    \caption{A three-layer feed forward neural network.}
    \label{fig:mlp}
\end{figure}

Artificial Neural Networks is an algorithm which can model non-linear relationships between input and output data \cite{Akaike1974}. A popular neural network is a feed-forward multilayer perceptron. Fig. \ref{fig:mlp} shows a three-layer feed-forward neural network with a single output unit, \textit{k} hidden units, $n$ input units. $w_{ij}$ is the connection weight from the $i^{th}$ input unit to the $j^{th}$ hidden unit,  and $T_j$ is the connecting weight from the $j^{th}$ hidden unit to the output unit \cite{Pao2007}. These weights transform the input variables in the first layer to the output variable in the final layer using the training data. 


For a univariate time series forecasting problem, suppose we have N observations $y_1, y_2, \ldots, y_N$ in the training set, and $m$ observations in the test set, $y_{N+1}, y_{N+2}, \ldots, y_{N+m}$, and we are required to predict \textit{m} periods ahead \cite{Pao2007}. The training patterns are as follows:
\begin{align}
y_{p+m} & =f_{W}(y_p, y_{p-1},\ldots,y_1)\\
y_{p+m+1} & =f_{W}(y_{p+1}, y_{p},\ldots,y_2)\\
&\vdotswithin  \notag \\
y_{N} & =f_{W}(y_{N-m},y_{N-m-1},\ldots,y_{N-m-p+1})
\end{align}

\noindent where $f_{W}(\cdot)$ represents the MLP network and $W$ are the weights. For brevity we omit $W$. The training patterns use previous time-series points, for example, $y_p, y_{p-1},\ldots,y_1$ as the time series is univariate. That is, we only have the time series in which we can draw inferences from. In addition, these time series points are correlated, and therefore provide information that can be used to predict the next time point.

The $m$ testing patterns are 

\begin{align}
y_{N+1} & =f_{W}(y_{N+1-m}, y_{N-m},\ldots,y_{N-m-p+2})\\
y_{N+2} & =f_{W}(y_{N+2-m}, y_{N-m+1},\ldots,y_{N-m-p+3})\\
&\vdotswithin  \notag \\
y_{N+m} & =f_{W}(y_{N},y_{N-1},\ldots,y_{N-p+1}).
\end{align}

The training objective is to minimize the overall predictive mean sum of squared estimate of errors (SSE) by adjusting the connection weights. For this network structure the SSE can be written as $\sum_{i=p+m}^N(y_i-\hat{y}_i)$ where $\hat{y}_i$ is the prediction from the network. The number of input nodes corresponds to the number of lagged observations. Having too few or too many input nodes can affect the predictive ability \cite{Pao2007}.


It is also possible to vary the the number of input units. Typically, various different configurations of units are trialled, with the best configuration being used in production. The weights $W$ in $f_W$ are trained using a process called backpropagation, which uses labelled data and gradient descent to optimize the weights.

\subsection{Online Algorithms}

\noindent{\textit{\textbf{Box-Cox regressor.}}} In this subsection, we discuss the Box-Cox regressor. Ordinary least square is a method for estimating the unknown parameters in a linear regression algorithm. It estimates these unknown parameters by the principle of least squares. Specifically, it minimizes the sum of the squares of the differences between the observed variables and those predicted by the linear function.

The ordinary least squares regression assumes a normal distribution of residuals. However, when this is not the case, the Box-Cox Regression may be useful \cite{Box1964}. It transforms the dependent variable using the Box-Cox Transformation function and employs maximum likelihood estimation to determine the optimal level of the power parameter lambda. 

\noindent{\textit{\textbf{Passive-Aggressive regressor.}}} The goal of the Passive-Aggressive (PA) algorithm is to change itself as little as possible to correct for any mistakes and low-confidence predictions it encounters \cite{Gzik2014}. Specifically, with each example PA solves the following optimisation \cite{Ma2009}:
\begin{align}
    \boldsymbol{w}_{t+1}\leftarrow argmin \frac{1}{2}\left|\left|{\boldsymbol{w}_t-\boldsymbol{w}}\right|\right|^2 
    s.t. \; \; y_i(\boldsymbol{w}\cdot \boldsymbol{x}_t)\geq1.
\end{align}

\noindent Where $x_t$ is the input data and $y_i$ the output data, and $w_t$ are the weights for the PA algorithm and $w$ is the weight to be optimised. Updates occur when the inner product does not exceed a fixed confidence margin - i.e., $y_i(\boldsymbol{w}\cdot \boldsymbol{x}_t)\geq1$. The closed-form update for all examples is as follows:
\begin{equation}
    \boldsymbol{w}_{t+1}\leftarrow \boldsymbol{w}_{t} + \alpha_t y_t \boldsymbol{x}_t
\end{equation}

\noindent where 
\begin{equation}
\alpha_t=max\left\{\frac{1-y_t(\boldsymbol{w}_t\cdot\boldsymbol{x}_t)}{\left|\left|\boldsymbol{x}_t\right|\right|^2},0\right\}. 	
\end{equation}
\noindent $a_t$ is derived from a derivation process which uses the Lagrange multiplier \cite{Gzik2014}.

\subsection{Long-term Energy Market Model}

In order to test the impact of the different residual distributions, we used the ElecSim simulation \cite{Kell,Kell2020}. ElecSim is an agent-based model which mimics the behaviour of decentralized electricity markets. In this paper, we parametrized the model with data of the United Kingdom in 2018. This enabled us to create a digital twin of the UK electricity market and project forward. The data used for this parametrization included power plants in operation in 2018 and the funds available to the GenCos \cite{dukes_511, companies_house}.

\begin{figure}
\centering
    \includegraphics[width=0.4\textwidth]{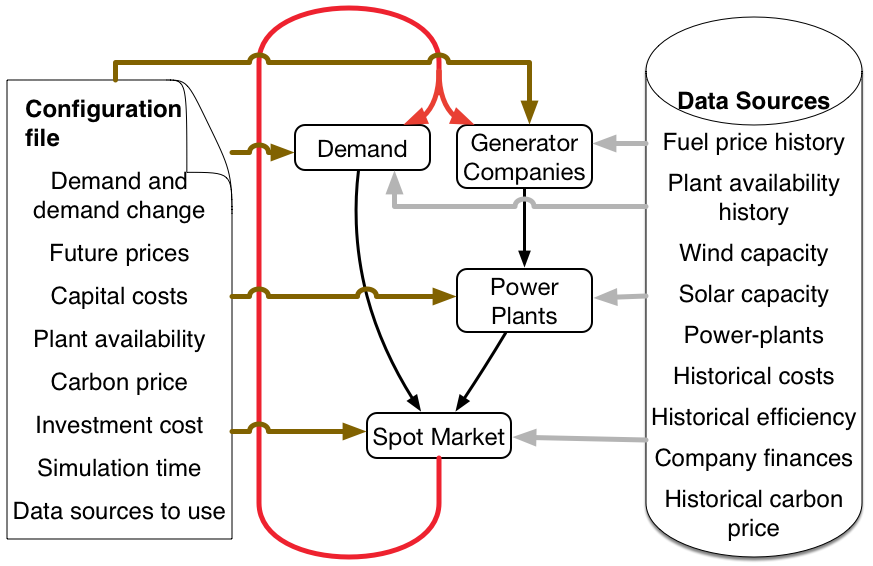}
    \caption{System overview of ElecSim \cite{Kell}.}
    \label{fig:model_details}
\end{figure}

ElecSim is made up of six components: 1) power plant data; 2) scenario data; 3) the time-steps of the algorithm; 4) the power exchange; 5) the investment algorithm and 6) the GenCos as agents. ElecSim uses a subset of representative days of electricity demand, solar irradiance and wind speed to approximate a full year. Representative days are a subset of days which, when scaled up, represent an entire year \cite{Kell2020}. We show how these components interact in Figure \ref{fig:model_details} \cite{Kell}. Namely, electricity demand is matched with the supply provided by power plants through the use of a spot market. Generator companies invest in power plants based upon information provided by the data sources and expectation of the data provided by the configuration file.

ElecSim uses a configuration file which details the scenario which can be set by the user. This includes electricity demand, carbon price and fuel prices. The data sources parametrize the ElecSim simulation to a particular country, including information such as wind capacity and power plants in operation. A spot market then matches electricity demand with supply.

The market runs a merit-order dispatch model, and bids are made by the power plant's short-run marginal cost (SRMC). A merit-order dispatch model is one which dispatches the cheapest electricity generators first. SRMC is the cost it takes to dispatch a single MWh of electricity and does not include capital costs. Investment in power plants is based upon a net present value (NPV) calculation. NPV is the difference between the present value of cash inflows and the present value of cash outflows over a period of time. This is shown in Equation \ref{eq:npv_eq}, where $t$ is the year of the cash flow, $i$ is the discount rate, $N$ is the total number of years, or lifetime of power plant, and $R_t$ is the net cash flow of the year $t$:
\begin{equation} \label{eq:npv_eq}
NPV(t, N) = \sum_{t=0}^{N}\frac{R_t}{(1+i)^t}.
\end{equation}
Each of the Generator Companies (GenCos) estimate the yearly income for each prospective power plant by running a merit-order dispatch electricity market simulation ten years into the future. However, it is true that the expected cost of electricity ten years into the future is particularly challenging to predict. We, therefore, use a reference scenario projected by the UK Government Department for Business and Industrial Strategy (BEIS), and use the predicted costs of electricity calibrated by Kell \textit{et al} \cite{Kell2020, DBEIS2019}. The agents predict the future carbon price by using a linear regression algorithm.

More concretely, the GenCos make investments by comparing the expected profitability of each of the power plants over their respective lifetime. They invest in the power plant which they deem to be the most profitable. A major uncertainty in power plant investment is the price of electricity whilst the power plant is in operation. This is often a 25 year period. This is where the predicted costs of electricity calibrated in \cite{Kell2020} is used.
 
\section{Methods}
\label{sec:methods}


\noindent{\textit{\textbf{Data Preparation.}}} Similarly to our previous work in \cite{Kell2018a}, we selected a number of calendar attributes and demand data from the GB National Grid Status dataset provided by the electricity market settlement company Elexon, and the University of Sheffield \cite{gbnationalgridstatus_2019}. This dataset contained data between the years 2011-2018 for the United Kingdom. The calendar attributes used as predictors to the algorithms were hour, month, day of the week, day of the month and year. These attributes allow us to account for the periodicity of the data within each day, month and year.

It is also the case that electricity demand on a public holiday which falls on a weekday is dissimilar to load behaviours of ordinary weekdays \cite{Kim2000}. We marked each holiday day to allow the algorithm to account for this.

As demand data is highly correlated with historical demand, we lagged the input demand data. In this context, the lagged data is where we provide data of previous time steps at the input. For example, for predicting $t+1$, we use $n$ inputs: $t,t-1,t-2,\ldots,t-n$. This enabled us to take into account correlations on previous days, weeks and the previous month. Specifically, we used the previous 28 hours before the time step to be predicted for the previous 1st, 2nd, 7th and 30th day. We chose this as we believe that the previous two days were the most relevant to the day to be predicted, as well as the weekday of the previous week and the previous month. We chose the previous 28 hours to account for a full day, plus an additional 4 hours to account for the previous day's correlation with the day to be predicted. We could have increased the number of days provided to the algorithm. However, due to time and computational constraints, we used our previously described intuition for lagged data selection. 

In addition to this, we marked each of the days with their respective seven seasons. These seasons were defined by the National Grid Short Term Operating Reserve (STOR) Market Information Report \cite{ESO2019}. These differ from the traditional four seasons by splitting autumn into two further seasons, and winter into three seasons. Finally, to predict a full 24-hours ahead, we used 24 different algorithms, 1 for each hour of the day.

The data is standardized and normalized using min-max scaling between -1 and 1 before training and predicting with the algorithm. This is due to the fact that the inputs such as day of the week, hour of day are significantly smaller than that of demand. Therefore, the demand will influence the result more due to its larger value. However, this does not necessarily mean that demand has greater predictive power.

\noindent{\textit{\textbf{Algorithm Tuning.}}} To find the optimum hyperparameters, cross-validation is used. As this time-series data were correlated in the time-domain, we took the first six years of data (2011-2017) for training and tested on the remaining year of data (2017-2018).

Each machine learning algorithm has a different set of parameters to tune. To tune the parameters in this paper, we used a grid search method. Grid search is a brute force approach that trials each combination of parameters at our choosing; however, for our search space was small enough to make other approaches not worth the additional effort.

Tables \ref{table:hyperparameter-tuning-offline} and \ref{table:hyperparameter-tuning-online} display each of the algorithms and respective parameters that were used in the grid search. Table \ref{table:hyperparameter-tuning-offline} shows the offline machine learning methods, whereas Table \ref{table:hyperparameter-tuning-online} displays the online machine learning methods. Each of the parameters within the columns ``Values'' are trialled with every other parameter.

Whilst there is room to increase the total number of parameters, due to the exponential nature of grid-search, we chose a smaller subset of hyperparameters, and a larger number of regressor types. Specifically, with neural networks, there is a possibility to extend the number of layers as well as the number of neurons, to use a technique called deep learning. Deep learning is a class of neural networks that use multiple layers to extract higher levels of features from the input. For this paper, however, we decided to trial a large number of different algorithms, instead of a large number of different configurations for neural networks.

\begin{sidewaystable}
\centering
\footnotesize
\begin{tabular}{@{}lllllll@{}}
\toprule
\textbf{Regressor Type} & \textbf{Parameters} & \textbf{Values}   & \textbf{Parameters} & \textbf{Values} & \textbf{Parameters} & \textbf{Values}       \\ \midrule
Linear, Lasso, Elastic Net, Least-Angle                  & N/A                 & N/A               &                     &                 &                     &                       \\
Extra Trees, Random Forest, AdaBoost             & \# Estimators       & {[}16, 32{]}      &                     &                 &                     &                       \\
Gradient Boosting       & \# Estimators       & {[}16, 32{]}      & learning rate       & {[}0.8, 1.0{]}  &                     &                       \\
Support Vector          & Kernel              & {[}linear, rbf{]} & C                   & {[}1, 10{]}     & Gamma               & {[}0.001, 0.0001{]}   \\
Multilayer Perceptron   & Activation function & {[}tanh, relu{]}  & hidden layer sizes  & {[}1, 50{]}     & Alpha               & {[}0.00005, 0.0005{]} \\
K-Neighbours            & \# Neighbours       & {[}5, 20, 50{]}   &                     &                 &                     &                       \\ \bottomrule
\end{tabular}%
\caption{Hyperparameters for offline machine learning regression algorithms}
\label{table:hyperparameter-tuning-offline}


\qquad
\qquad
\qquad
\qquad

\centering
\begin{tabular}{@{}llp{3cm}lllp{1.6cm}@{}}
\toprule
\textbf{Regressor Type} & \textbf{Parameters} & \textbf{Values}                                  & \textbf{Parameters} & \textbf{Values}   & \textbf{Parameters} & \textbf{Values}        \\ \midrule
Linear                  & N/A                 & N/A                                              &                     &                   &                     &                        \\
Box-Cox                 & Power               & {[}0.1, 0.05, 0.01{]}                            &                     &                   &                     &                        \\
Multilayer Perceptron   & Hidden layer sizes  & {[}(10, 50, 100), (10),  (20), (50), (10, 50){]} & 
                    &                   &                     &                        \\ 
                    Passive Aggressive      & C                   & {[}0.1, 1, 2{]}                                  & Fit intercept?      & {[}True, False{]} & Max iterations      & {[}1, 10, 100, 1000{]} \\
\bottomrule
\end{tabular}%
\caption{Hyperparameters for online machine learning regression algorithms}
\label{table:hyperparameter-tuning-online}
\end{sidewaystable}%

\noindent{\textit{\textbf{Implementation methodology.}}} The implementation slightly differs between online and offline learning. In offline learning, batch processing occurs. That is, all the data from 2011 to 2017 is used to train the algorithm. Once each of the algorithms had been trained, the algorithms are used to predict the electricity demand from 2017 to 2018. For hyper-parameter tuning cross-validation is used. Specifically, the training data is randomly split ten times to select the best hyper-parameters. This allowed for an unbiased assessment of the data.  

For online learning a similar process is undertaken. That is, the algorithms are trained in a batched approach using data from 2011 to 2017. Between the year 2017 and 2018, the next time-step is predicted and the error recorded between actual value and the predicted value. Next, the algorithm is updated using the actual value, and the next value predicted again.

\noindent{\textit{\textbf{Prediction Residuals in ElecSim.}}} Each of the algorithms trialled will have a degree of error. Prediction residuals are the difference between the estimated and actual values. We collect the prediction residuals to form a distribution for each of the algorithms. We then trial 80 different closed-form distributions to see which of the distributions best fits the residuals from each of the algorithms. These 80 distributions were chosen due to their implementation in scikit-learn \cite{scikit-learn}.

Once each of the prediction residual distributions are fit with a sensible closed-form distribution, we sample from this new distribution and perturb the demand for the electricity market at each time step within ElecSim.

By perturbing the market by the residuals, we can observe what the effects are of incorrect predictions of demand in an electricity market using ElecSim. 

ElecSim has previously been validated in \cite{Kell2020}. In this work, we validated our algorithm between the years 2013 and 2018, and recorded the difference between observed electricity mix to predicted electricity mix. We found that we were able to predict each different electricity source better than the naive approach. The naive approach, in this case, was predicting the electricity mix at the last known point in 2013. However, we were able to better predict coal, solar and wind by achieving a mean absolute scaled error (MASE) of ${\sim}$0.4 for these. CCGT and nuclear on the other hand had slightly worse results, achieving a MASE of ${\sim}$0.7.

\section{Results}
\label{sec:results}

In this Section, we detail the accuracy of the algorithms to predict 24 hours ahead for the day-ahead market. In addition to this, we display the impact of the errors on electricity generation investment and electricity mix from the years 2018 to 2035 using the agent-based model ElecSim.

\subsection{Offline Machine Learning}

In this work, the training data was from 2011 to 2017, and the testing data was from 2017 to 2018.

\begin{figure}[h]
\centering
\includegraphics[width=\columnwidth]{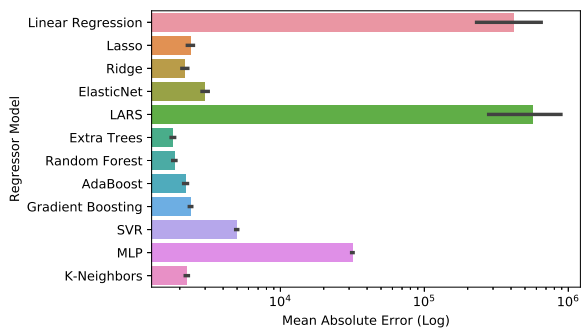}
\caption{Offline algorithms mean absolute error comparison, with 95\% confidence interval for 5 runs of each algorithm.}
\label{fig:beis_elecsim_historic_comparison}
\end{figure}

Figure \ref{fig:beis_elecsim_historic_comparison} displays the mean absolute error of each of the offline statistical and machine learning algorithms on a log scale. It can be seen that the different algorithms have varying degrees of success. The least accurate algorithms were linear regression, multilayer perceptron (MLP) and the Least Angle Regression (LARS), each with mean absolute errors over 10,000MWh. This error would be prohibitively high in practice; the max tendered national grid reserve is 6,000MWh, while the average tendered national grid reserve is 2,000MWh \cite{ESO2019}.

A number of algorithms perform well, with a low mean absolute error. These include the Lasso, gradient Boosting Regressor and K-neighbours regressor. The best algorithm, similar to \cite{Kell2018a}, was the decision tree-based algorithm, Extra Trees Regressor, with a mean absolute error of $1,604$MWh. This level is well within the average national grid reserve of 2,000MWh.

Table \ref{table:offline_ml_metrics} displays different metrics for measuring the accuracy of the offline machine learning techniques. These include Mean Squared Error (MSE), Root Mean Squared Error (RMSE), Mean Absolute Error (MAE) and the Mean R-Squared. The results largely compliment each other. With high error values in one metric correlating to high error metrics in others for each of the estimators.

\begin{table*}[]
\footnotesize
	\begin{tabular}{@{}lllllll@{}}
		\toprule
		Estimator                 & Mean Fit Time & Mean Score Time & Mean MSE    & Mean RMSE & Mean MAE & Mean R-Squared \\ \midrule
		LinearRegression          & 57            & 4.84            & 9444808.95  & 3073.24   & 2249.34  & 0.81           \\
		Lasso                     & 787.97        & 3.13            & 9446957.22  & 3073.59   & 2249.55  & 0.81           \\
		Ridge                     & 26.99         & 5.03            & 9444701.58  & 3073.22   & 2252.23  & 0.81           \\
		ElasticNet                & 54.49         & 6.13            & 31139231.96 & 5580.25   & 4628.64  & 0.36           \\
		llars                     & 46.85         & 8               & 10164815.45 & 3188.23   & 2333.31  & 0.79           \\
		ExtraTreesRegressor       & 9321.52       & 58.06           & 5562579.53  & 2358.51   & 1605     & 0.89           \\
		RandomForestRegressor     & 16567.46      & 13.99           & 5882618.89  & 2425.41   & 1646.29  & 0.88           \\
		AdaBoostRegressor         & 8897.55       & 26.9            & 18551963.36 & 4307.2    & 3544.49  & 0.62           \\
		GradientBoostingRegressor & 6417.62       & 8.16            & 6744402.87  & 2597      & 1833.62  & 0.86           \\
		SVR                       & 19170.82      & 5221.66         & 51217167.5  & 7156.62   & 5926.75  & -0.05          \\
		KNeighborsRegressor       & 118.89        & 15215.87        & 10107201.23 & 3179.18   & 2246.6   & 0.79           \\ \bottomrule
	\end{tabular}
\caption{Different metrics for offline machine learning results.}
\label{table:offline_ml_metrics}
\end{table*}

\begin{figure}[h]
\centering
\includegraphics[width=\columnwidth]{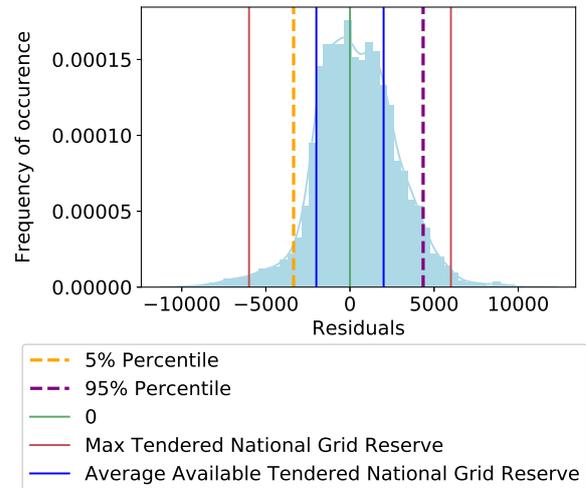}
\caption{Best offline machine learning algorithm (Extra Trees Regressor) distribution.}
\label{fig:best_offline_learning_day_distribution}
\end{figure}

Figure \ref{fig:best_offline_learning_day_distribution} displays the distribution of the best offline machine result (Extra Trees Regressor). It can be seen that the max tendered national grid reserve falls well above the 5\% and 95\% percentiles. However, there are occasions where the errors are greater than the maximum tendered national grid reserve. In addition, the majority of the time, the algorithm's predictions fall within the average available tendered national grid reserve. Therefore, there is room for improvement, to ensure that blackouts do not occur and predictions fall within the max tendered national grid reserve.

%
%

\begin{figure}[h]
\centering
\includegraphics[width=\columnwidth]{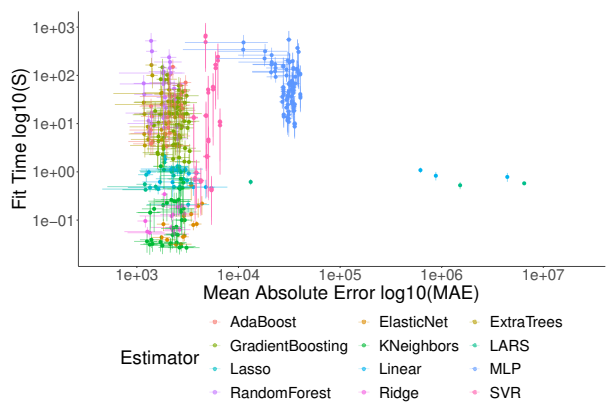}
\caption{Time taken to train the offline algorithms versus mean absolute error. Error bars display standard deviation between points.}
\label{fig:offline_fit_time_vs_mae}
\end{figure}

\begin{figure}[h]
\centering
\includegraphics[width=\columnwidth]{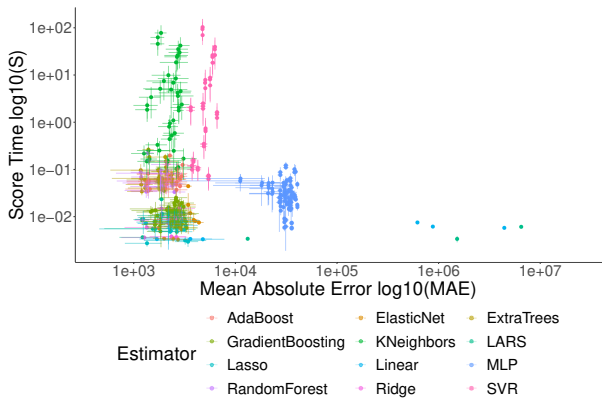}
\caption{Time taken to score the offline algorithm versus mean absolute error. Error bars display standard deviation between points.}
\label{fig:offline_score_time_vs_mae}
\end{figure}

Figures \ref{fig:offline_fit_time_vs_mae} and \ref{fig:offline_score_time_vs_mae} display the time taken to train the algorithm and time taken to sample from the algorithm versus the absolute error respectively for the offline algorithms. Multiple fits are trialled for each parameter type for each algorithm. The error bars indicate the results of multiple cross-validations.

It can be seen from Figure \ref{fig:offline_fit_time_vs_mae} that the time to fit varies significantly between algorithms and parameter choices. The multilayer perceptron consistently takes a long time to fit, when compared to the other algorithms and performs relatively poorly in terms of MAE. There are many algorithms, such as the random forest regressor, and extra trees regressors which perform well, however, take a long time to fit, especially when compared to the K-Nearest neighbours.

For a small deterioration in MAE it is possible to decrease the time it takes to train the algorithm significantly. For example, by using the K-Nearest neighbours or support vector regression (SVR).


The scoring time, displayed in Figure \ref{fig:offline_score_time_vs_mae}, also displays a large variation between algorithm types. For instance, the MLP regressor takes a shorter time to sample predictions when compared to the K-Neighbors algorithm and support vector regression. It is possible to have a cluster of algorithms with low sample times and low mean absolute errors. However, often a trade-off is required, with a fast prediction time requiring a longer training time and vice-versa. Practitioners, therefore, must decide which aspect is most important for them for their use case: speed of training or of prediction.

\subsection{Online Machine Learning}

\begin{figure}[h]
\centering
\includegraphics[width=0.8\columnwidth]{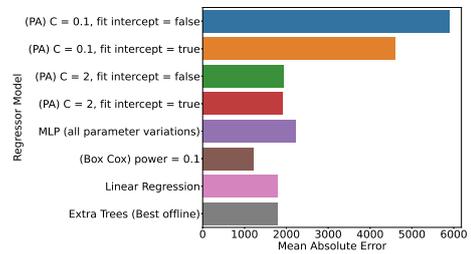}
\caption{Comparison of mean absolute errors (MAE) for different online regressor algorithms. MLP results for all parameters are shown in a single barchart due to the very similar MAEs for the differing hyperparameters.}
\label{fig:online_model_mae_barplot}
\end{figure}

To see if we can improve on the predictions, we utilize an online machine learning approach. If we are successful, we should be able to reduce the national grid reserves, reducing cost and emissions.

Figure \ref{fig:online_model_mae_barplot} displays the comparison of mean absolute errors for the different trialled online regressor algorithms. To produce this graph, we show various hyperparameter trials. Where the hyperparameters had the same results, we removed them. For the multilayer perceptron (MLP), we aggregated all hyperparameters, due to the similar nature of the predictions.

It can be seen that the best performing algorithm was the Box-Cox regressor, with an MAE of 1100. This is an improvement of over 30\% on the best offline algorithm. The other algorithms perform less well. However, it can be seen that the linear regression algorithm improves significantly for the online case when compared to the offline case. The passive aggressive (PA) algorithm improve significantly with the varying parameters, and the MLP performs poorly in all cases.

Table \ref{table:online_ml_metrics} displays the metrics for each of the online methods. This includes the mean MSE, mean RMSE and mean MAE. Again, the metrics largely correlate with each other. Meaning, that the favoured metric can be used when selecting an estimator.

%

\begin{table*}[]
\centering
\footnotesize
\begin{tabular}{@{}llllll@{}}
\toprule
Estimator                                                                          & Mean Fit Time & Mean Score Time & Mean MSE     & Mean RMSE & Mean MAE \\ \midrule
(PA) C = 0.1, fit intercept = false & 173.52        & 24.63           & 103015609.55 & 9497.47   & 5888.4   \\
(PA) C = 0.1, fit intercept = true     & 168.66        & 24.07           & 63201775.28  & 7430.63   & 4605.94  \\
(PA) C = 2, fit intercept = false      & 165.23        & 23.75           & 7451087.49   & 2723.25   & 1927.33  \\
(PA) C = 2, fit intercept = true       & 174.91        & 24.65           & 7223163.14   & 2681.2    & 1907.59  \\
(MLP) (all parameter variations)                                                     & 71.36         & 6.58            & 9612351.37   & 3076.77   & 2221.48  \\
(Box Cox) power = 0.1                                                              & 38.61         & 4.88            & 2921934.52   & 1703.79   & 1214.95  \\
Linear Regression                                                                  & 38.61         & 4.85            & 5629651.14   & 2368.3    & 1785.02  \\ \bottomrule
\end{tabular}
	\caption{Different metrics for online machine learning results.}
	\label{table:online_ml_metrics}
\end{table*}

\begin{figure}[h]
\centering
\includegraphics[width=0.8\columnwidth]{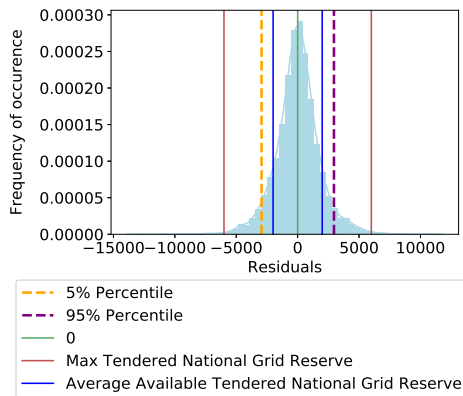}
\caption{Best online algorithm (Box-Cox Regressor) distribution.}
\label{fig:best_online_learning_day_distribution}
\end{figure}

Figure \ref{fig:best_online_learning_day_distribution} displays the best online algorithm. We can see a significant improvement over the best online algorithm distribution, shown in Figure \ref{fig:best_offline_learning_day_distribution}. We remain within the max tendered national grid reserve for 98.9\% of the time, and the average available tendered national grid reserve is close to the 5\% and 95\% percentiles. This model, therefore would be highly recommended over the best offline version to ensure demand is matched with supply.

\begin{figure}[h]
\centering
\includegraphics[width=0.8\columnwidth]{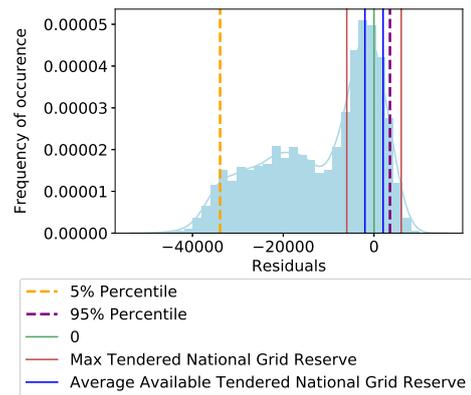}
\caption{Online machine learning algorithm distribution. (Passive Aggressive Regressor (C=0.1, fit intercept = true, maximum iterations = 1000, shuffle = false, tolerance = 0.001), chosen as it was the worst result for the passive aggressive algorithm.}
\label{fig:bad_online_learning_day_distribution}
\end{figure}

Figure \ref{fig:bad_online_learning_day_distribution} displays the residuals for a algorithm with poor predictive ability, the passive aggressive regressor. It displays a large period of time of prediction errors at -20,000MWh, and often falls outside of the national grid reserve. These results demonstrate the importance of trying a multitude of different algorithms and parameters to improve prediction accuracy.

\begin{figure}[h]
\centering
\includegraphics[width=0.8\columnwidth]{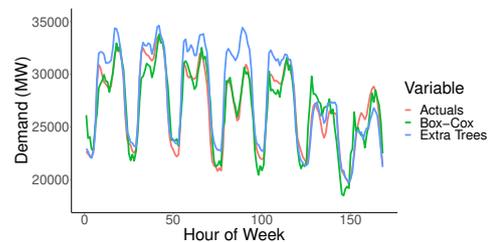}
\caption{Best offline algorithm compared to the best online algorithm over a one week period.}
\label{fig:both_actual_predicted}
\end{figure}

Figure \ref{fig:both_actual_predicted} displays a comparison between the actual electricity consumption compared to the predictions. It can be seen that the Box-Cox algorithm better predicts the actual electricity demand in most cases when compared to the best offline algorithm, the Extra Trees regressor. The Extra Trees regressor often overestimates the demand, particularly during weekdays. Whilst the Box-Cox regressor more closely matches the actual results. During the weekend (between the hours of 120 and 168), the Extra Trees regressor performs better, particularly on the Saturday (between hours of 144 and 168).

\begin{figure}[h]
\centering
\includegraphics[width=\columnwidth]{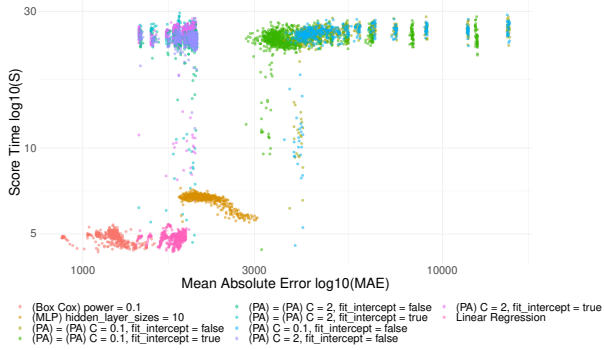}
\caption{Time taken to test the online algorithms versus mean absolute error.}
\label{fig:online_test_vs_mae}
\end{figure}

\begin{figure}[h]
\centering
\includegraphics[width=\columnwidth]{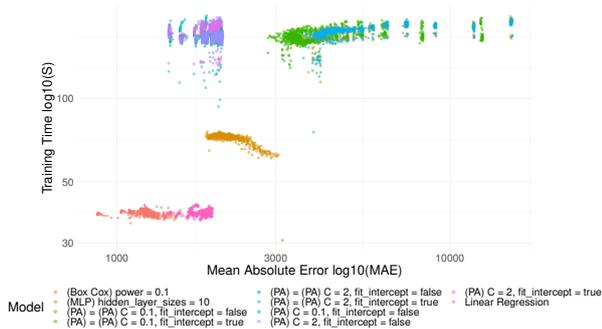}
\caption{Time taken to train the online algorithms versus mean absolute error.}
\label{fig:online_train_vs_mae}
\end{figure}

Figures \ref{fig:online_test_vs_mae} and \ref{fig:online_train_vs_mae} display the mean absolute error versus test and training time respectively. In these graphs, a selection of algorithms and parameter combinations are chosen. 

Clear clusters can be seen between different types of algorithms and parameter types. With the passive aggressive (PA) algorithm performing the slowest for both training and testing. Different parameter combinations show different results in terms of mean absolute error.

The best performing algorithm is the Box-Cox algorithm, which is also the fastest to both train and test. The linear regression, which performs worse in terms of predictive performance, is as quick to train and test as the Box-Cox algorithm. Additionally, the multilayer perceptron (MLP) is relatively quick to train and test when compared to the PA algorithms. We, therefore, recommend the Box-Cox algorithm as an optimal for training and testing speed as well as accuracy.

It is noted that when compared to the offline algorithms, the training time is a good indicator to the testing time. In other words, algorithms that are fast to train are also fast to test and vice-versa.

\subsection{Scenario Comparison}

In this Section we explore the effect of these residuals on investments made and the electricity generation mix.  To generate these graphs, we perturbed the exogenous demand in ElecSim by sampling from the best-fitting distributions for the respective residuals of each of the online methods. We did this for all of the online learning algorithms displayed in Figure \ref{fig:online_model_mae_barplot}. We let the simulation run for 17 years from 2018 to 2035. 

Running this simulation enabled us to see the effect on carbon emissions on the electricity grid over a long time period. For instance, does underestimating electricity demand mean that peaker power plants, such as reciprocal gas engines, are over utilized when other, less polluting power plants could be used?

\subsubsection{Mean Contributed Energy Generation}


\begin{figure*}[h]
\includegraphics[width=\textwidth]{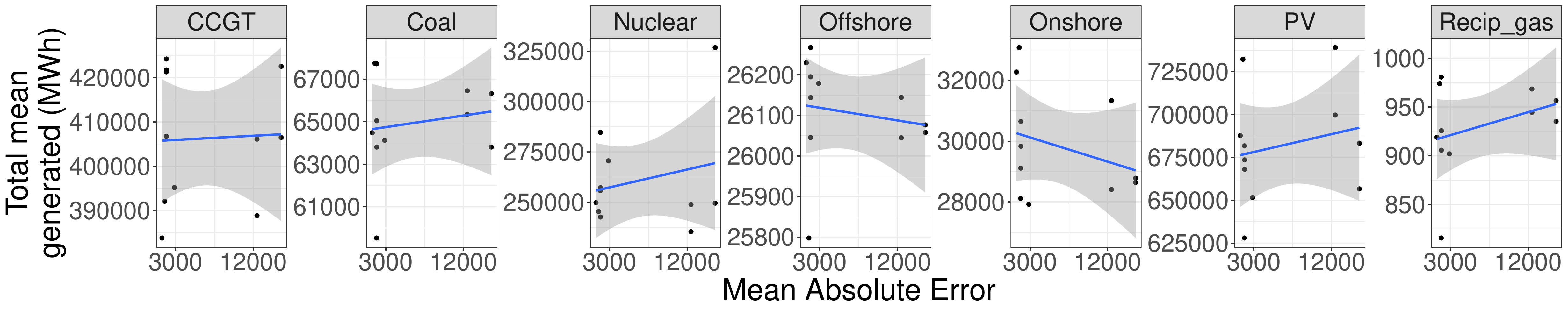}
\caption{Mean outputs of various technologies vs. mean absolute error from 2018 to 2035 in ElecSim.}
\label{fig:pv_coal_nuclear_offshore_outputs}
\end{figure*}

In this Section we display the mean electricity mix contributed by different electricity sources from 2018 to 2035. 

Figure \ref{fig:pv_coal_nuclear_offshore_outputs} displays the mean power contributed between 2018 and 2035 for each source vs. mean absolute error of the various online regressor algorithms displayed in Figure \ref{fig:online_model_mae_barplot}. A positive correlation can be seen with PV contributed and mean absolute error. This is similar for coal and nuclear output. However, it can be seen that offshore wind reduces with mean absolute error. Output for the reciprocal gas engine also increases with mean absolute error.

The reciprocal gas engine was expected to increase with times of high error. This is because, traditionally, reciprocal gas engines are peaker power plants. Peaker power plants provide power at times of peak demand, which cannot be covered by other plants due to them being at their maximum capacity level or out of service. It may also be the case, that with higher proportions of intermittent technologies, there is a larger need for these peaker power plants to fill in for times where there is a deficit in wind speed and solar irradiance.

It is hypothesized that coal and nuclear output increase to cover the predicted increased demands of the service. As these generation types are dispatchable, meaning operators can choose when they generate electricity, they are more likely to be used in times of higher predicted demand.

\subsubsection{Total Energy Generation}

In this Section, we detail the difference in total technologies invested in over the time period between 2018 to 2035, as predicted by ElecSim.

\begin{figure*}[h]
\includegraphics[width=\textwidth]{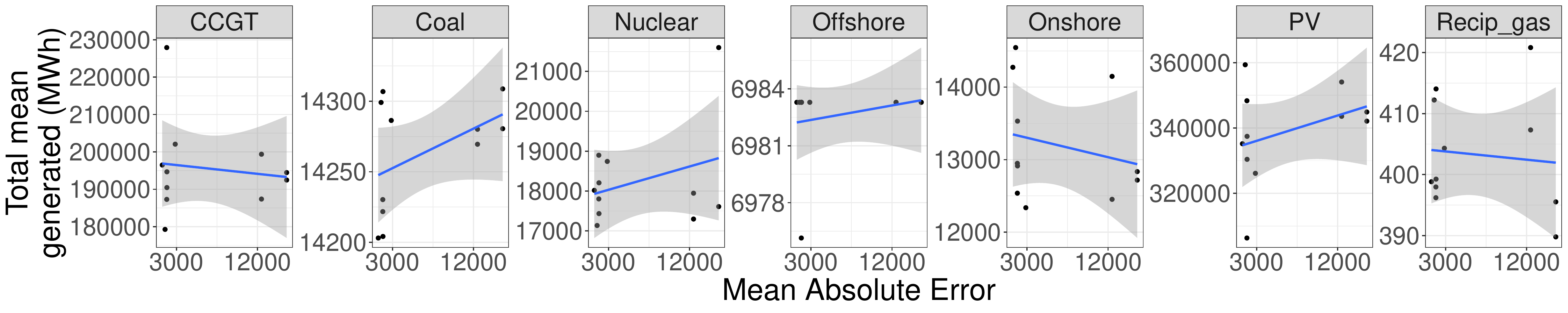}
\caption{Total technologies invested in vs. mean absolute error from 2018 to 2035 in ElecSim.}
\label{fig:ccgt_coal_onshore_offshore_totals}
\end{figure*}

CCGT, onshore, and reciprocal gas engines investment is less with an increase in MAE, as shown in Figure \ref{fig:ccgt_coal_onshore_offshore_totals}. While coal, offshore, nuclear and photovoltaics all exhibit increasing investments with MAE. Therefore, a smaller error leads to an increased usage of onshore wind, where lulls in wind supply are covered by CCGT and reciprocal gas engines.

It is hypothesized that coal and nuclear increase in investment due to their dispatchable nature. While onshore, non-dispatchable by nature, become a less attractive investment.

CCGT and reciprocal gas engines may have decreased in capacity over this time, due to the increase in coal. This could be because of the large consistent errors in prediction accuracy that meant that reciprocal gas engines were perceived to be less valuable.


\begin{figure}
\centering
\includegraphics[width=0.6\columnwidth]{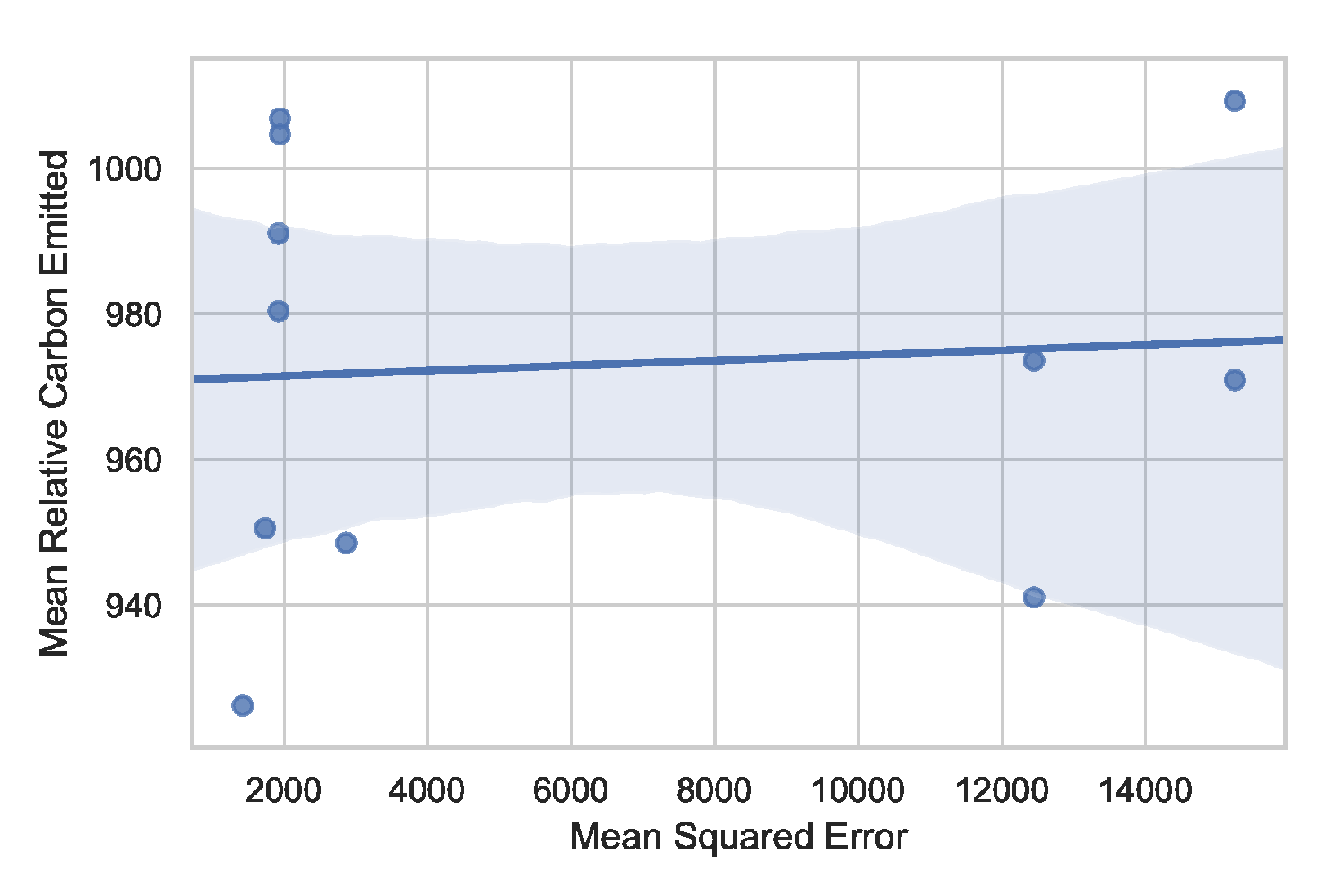}
\caption{Mean carbon emissions between 2018 and 2035.}
\label{fig:Carbon_emitted_mean_output}
\end{figure}

Figure \ref{fig:Carbon_emitted_mean_output} shows an increase in relative mean carbon emitted with mean absolute error of the predictions residuals. The reason for an increase in relative carbon emitted could be due to the increased output of utility of the reciprocal gas engine, coal, and decrease in offshore output. Reciprocal gas engines are peaker plants and, along with coal, can be dispatched. By being dispatched, the errors in predictions of demand can be filled. It is therefore recommended that by improving the demand prediction algorithms, significant gains can be made in reducing carbon emissions.


\subsubsection{Sensitivity Analysis}

In this Section we run a sensitivity analysis to visualise the effects of different errors on the average electricity mix over the 2018 to 2035 time period. To conduct this sensitivity analysis, we used a normal distribution with a mean of 0 and modified the standard deviation between 1,000 and 20,000, in increments of 1,000. We selected the normal distribution due to its observation in nature, and its symmetric properties. We chose to increase the standard deviation until 20,000 due to it being 33\% larger than the errors shown in the previous subsection. This gave us a larger error than had previously been explored.

\begin{figure}[h]
\centering
\includegraphics[width=0.7\columnwidth]{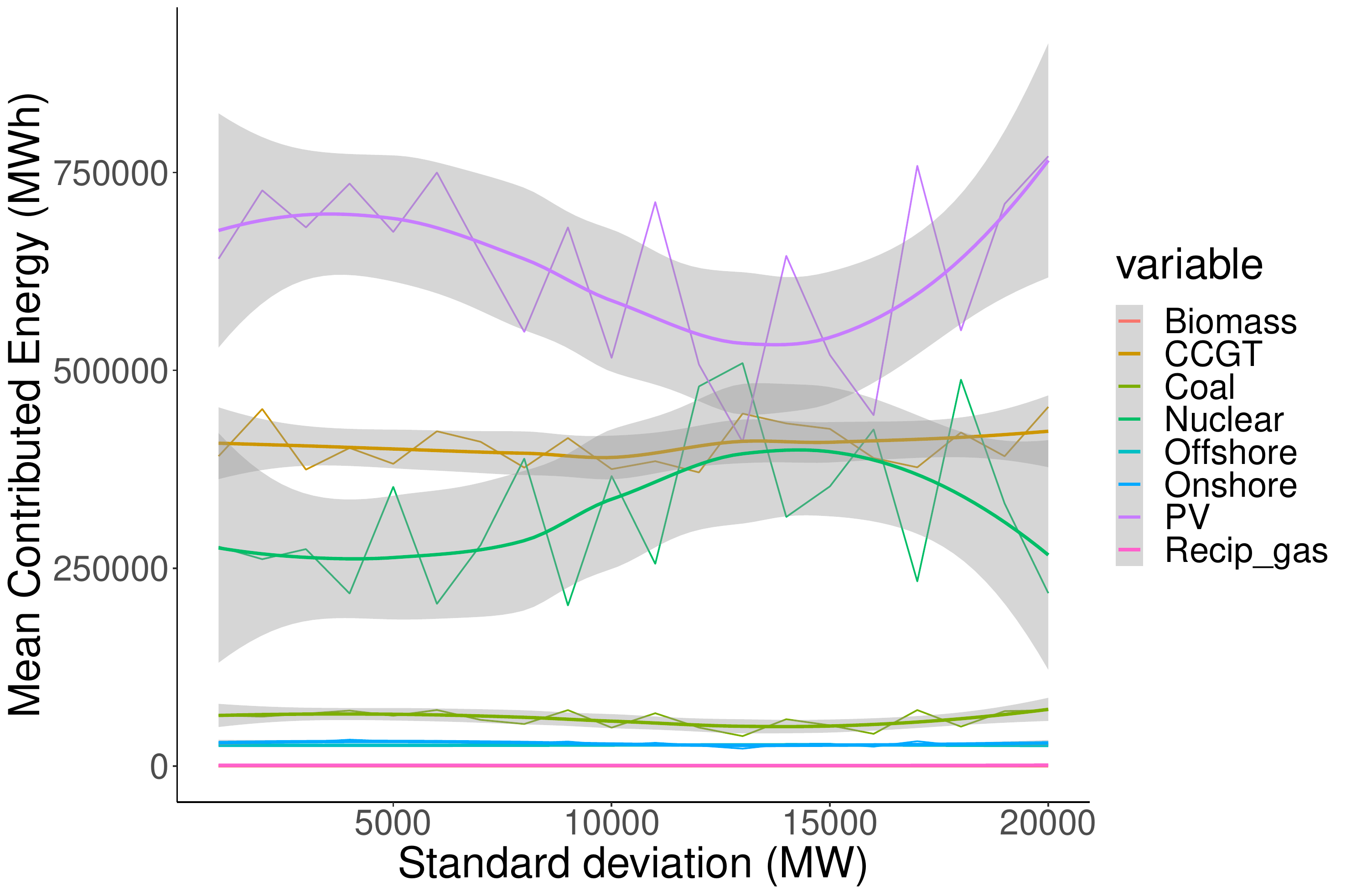}
\caption{Sensitivity analysis of changing demand prediction error using a normal distribution and varying the standard deviation vs. mean contributed energy per type between 2018 and 2035.}
\label{fig:sensitivity-analysis}
\end{figure}

Figure \ref{fig:sensitivity-analysis} displays the results of this sensitivity analysis. At all standard deviations, photovoltaics displays the highest contributed energy, CCGT the second most and nuclear the third most. Coal, onshore and the others contribute less energy than the top three.

Whilst photovoltaics remains high throughout, it reduces up until a standard deviation of 14,000MW, after which it begins to increase. This reduction up until 14,000MW may be due to the fact that the predicted demand changes so quickly, that photovoltaics are unable to be reliably dispatched on the day-ahead market to meet this changing predicted demand. Nuclear is, however, able to fill the demand that photovoltaics can not, due to its dispatchable nature.

After a standard deviation of 14,000MW, photovoltaics increases, whilst nuclear decreases. This may be due to the large positive error predictions, which photovoltaics believes it is able to be dispatched on and displaces the more expensive nuclear energy technology.





\section{Discussion}
\label{sec:discussion}


From our results, it can be seen that different algorithms yield differing prediction accuracies. Online algorithms can result in a decrease in 30\% of prediction error on the best offline algorithms. We calculated this by comparing the MAE for Extra Trees to the MAE for the Box-Cox regressor. We, therefore, recommend the use of online machine learning for predicting electricity demand in a day-ahead market.

Similar to our assumptions, the online learning algorithms were able to outperform the offline algorithms. This is due to the non-stationary nature of the data. An online method is able to use the most up-to-date knowledge of the complex system of energy demand. For instance, a certain year may have a particularly warm winter when compared to previous years, reducing the amount of electricity used for heating.

However, contrary to our assumptions, the online linear regression techniques outperformed the online machine learning techniques. This may be due to their simpler nature and ability to learn from a smaller subset of new data as opposed to relying on a large historic subset. For the offline algorithms, the best performing algorithms were the decision tree approaches such as extra trees and random forests. This is a similar outcome to our previous work, which showed that the best performing method for demand forecasting were random forests \cite{Kell2018}. Contrary to our assumptions, however, the lasso and ridge regression outperformed the machine learning techniques support vector regression and multilayer perceptron. This may be due to the ability of feature selection by lasso and ridge regression, which only uses the most important features.

To the best of our knowledge, more work has been done using offline learning to predict electricity demand. This may be due to the additional complexity of running online algorithms, and a smaller number of available algorithms to run in an online fashion.

In terms of computing power, finding the optimal input parameters, hyperparameters and algorithms to use can be a large undertaking. This is due to the exponential growth of the number of choices that can be made. This can be an issue where accuracy is of importance, especially when the data changes over time, meaning it may be necessary to retest previous results. However, due to the financial and sustainability implications, we believe the trade-off between compute time and accuracy is balanced towards compute time. There are also large implications if the algorithm were to break at a certain point in time. We, therefore, recommend the reliance on multiple well-performing algorithms, as opposed to solely the best performing algorithm at any one time. 

For training time and prediction time, there is often a trade-off between training and predicting. For instance, the k-nearest neighbours is fast to train, but slow to sample from. Therefore stakeholders must make a decision based upon accuracy, speed of training and sampling. 

The amount of time taken to test and cross-validate the algorithms increases exponentially with the number of algorithms and hyper-parameters trialled. It is, therefore, suggested that cloud computing is used to train the algorithms. This would enable the trialling of many different algorithms and hyper-parameters within a reasonable time limit for this time-sensitive application. However, once the algorithms have been trained and are used for making predictions the predictions can be made within a two minutes in the worst case. For the application of predicting 24-hours ahead, this falls within a reasonable time.

The impact on the broader electricity market has been shown to be significant. Principally, the investment behaviours of GenCos change as well as the dispatched electricity mix. The relative mean carbon emitted over this time period increases, due to an increase in the utilization of coal and reciprocal gas engines, at the expense of offshore wind.

\section{Conclusion}
\label{sec:conclusion}

In this paper, we evaluated 16 different machine learning and statistical algorithms to predict electricity demand in the UK for the day-ahead market. Specifically, we used both online and offline algorithms to predict electricity demand 24 hours ahead. We compared the ability for the offline algorithms: lasso regression, random forests, support vector regression, for both online and offline learning: linear regression, multilayer perceptron and for just online learning: the Box-Cox transformation and the passive aggressive regressor, amongst others. The Box-Cox, as well as the passive aggressive regressors, were used as online learning algorithms, the multilayer perceptron and linear regression were used as both, whereas the rest were used as offline learning algorithms.

We measured the errors and compared these to each algorithm as well as the national grid reserve. We found that through the use of an online learning approach, we were able to significantly reduce error by 30\% on the best offline algorithm.  We were also able to reduce our errors to significantly below the national grid's mean and maximum tendered reserve, thus significantly reducing the chances of blackouts.

In addition to this, we took these errors, or residuals, and perturbed the electricity market of the agent-based model ElecSim. This enabled us to see the impact of different error distributions on the long-term electricity market, in terms of investment and electricity mix.

We observed that with an increase in prediction errors, we get a higher proportion of electricity generated by coal, offshore, nuclear, reciprocal gas engines and photovoltaics. This could be due to the fact that more peaker and dispatchable plants are required to fill in for unexpected demand. In addition, a higher proportion of intermittent renewable energy sources leads to a higher use of peaker power plants to fill in the gaps of intermittency of wind and solar irradiance. However, by reducing the mean absolute error, we are able to reduce the amount of reciprocal gas engine and coal usage.

In future work, we would like to trial a different selection of algorithms inputs to the algorithms, for instance, by providing the algorithm with two months worth of historical data as dependent variables. Additionally, we would like to see the impact of predicting wind speed and solar irradiance to see how these impact the overall investment patterns and electricity mix. We would also like to use ensemble models in future, which combine the results of multiple algorithms.

\section{Funding Sources}

This work was supported by the Engineering and Physical Sciences Research Council, Centre for Doctoral Training in Cloud Computing for Big Data [grant number EP/L015358/1].



  \bibliographystyle{elsarticle-num} 
  \bibliography{library,bib_custom,custombibtex}

\begin{thebibliography}{10}
\expandafter\ifx\csname url\endcsname\relax
  \def\url#1{\texttt{#1}}\fi
\expandafter\ifx\csname urlprefix\endcsname\relax\def\urlprefix{URL }\fi
\expandafter\ifx\csname href\endcsname\relax
  \def\href#1#2{#2} \def\path#1{#1}\fi

\bibitem{Lu1993}
C.~Lu, H.~T. Wu, S.~Vemuri, {Neural network based short term load forecasting},
  IEEE Transactions on Power Systems 8~(1) (1993) 336--342.
\newblock \href {https://doi.org/10.1080/02533839.1995.9677697}
  {\path{doi:10.1080/02533839.1995.9677697}}.

\bibitem{Kell2018a}
A.~Kell, A.~McGough, M.~Forshaw, {Segmenting residential smart meter data for
  short-Term load forecasting}, in: e-Energy 2018 - Proceedings of the 9th ACM
  International Conference on Future Energy Systems, 2018.

\bibitem{Hong2014}
T.~Hong, J.~Wilson, J.~Xie, A.~Member, {Long Term Probabilistic Load
  Forecasting and Normalization With Hourly Information} 5~(1) (2014) 456--462.

\bibitem{Al-Musaylh2018}
M.~Al-Musaylh, R.~Deo, J.~Adamowski, Y.~Li, {Short-term electricity demand
  forecasting with MARS, SVR and ARIMA models using aggregated demand data in
  Queensland, Australia}, Advanced Engineering Informatics 35~(November 2017)
  (2018) 1--16.

\bibitem{Vrablecova2017}
P.~Vrablecov{\'{a}}, A.~{Bou Ezzeddine}, V.~Rozinajov{\'{a}},
  S.~{\v{S}}{\'{a}}rik, A.~K. Sangaiah, {Smart grid load forecasting using
  online support vector regression}, Computers {\&} Electrical Engineering 0
  (2017) 1--16.
\newblock \href {https://doi.org/10.1016/j.compeleceng.2017.07.006}
  {\path{doi:10.1016/j.compeleceng.2017.07.006}}.

\bibitem{Huang2003}
S.-j. Huang, S.~Member, K.-r. Shih, {Short-Term Load Forecasting Via ARMA Model
  Identification Including Non-Gaussian} 18~(2) (2003) 673--679.

\bibitem{Andersen2013}
F.~M. Andersen, H.~V. Larsen, T.~K. Boomsma, {Long-term forecasting of hourly
  electricity load: Identification of consumption profiles and segmentation of
  customers}, Energy Conversion and Management 68 (2013) 244--252.

\bibitem{Kell}
A.~Kell, M.~Forshaw, A.~S. McGough, {ElecSim : Monte-Carlo Open-Source
  Agent-Based Model to Inform Policy for Long-Term Electricity Planning}, The
  Tenth ACM International Conference on Future Energy Systems (ACM e-Energy
  `19) (2019) 556--565.

\bibitem{Kell2020}
A.~J.~M. Kell, M.~Forshaw, A.~S. McGough, {Long-Term Electricity Market Agent
  Based Model Validation using Genetic Algorithm based Optimization}, The
  Eleventh ACM International Conference on Future Energy Systems (e-Energy'20)
  (2020).

\bibitem{Singh2012}
A.~K. Singh, Ibraheem, S.~Khatoon, M.~Muazzam, D.~K. Chaturvedi, {Load
  forecasting techniques and methodologies: A review}, ICPCES 2012 - 2012 2nd
  International Conference on Power, Control and Embedded Systems (2012).

\bibitem{Kell2018}
A.~Kell, A.~S. Mcgough, M.~Forshaw, {Segmenting Residential Smart Meter Data
  for Short-Term Load Forecasting}, e-Energy Conference (2018) 91--96.

\bibitem{Ahmad2017}
M.~W. Ahmad, M.~Mourshed, Y.~Rezgui, {Trees vs Neurons: Comparison between
  random forest and ANN for high-resolution prediction of building energy
  consumption}, Energy and Buildings 147 (2017) 77--89.

\bibitem{Chen2004}
B.-j. Chen, M.-w. Chang, C.-j. Lin, {Load Forecasting Using Support Vector
  Machines : A Study on EUNITE Competition 2001}, IEEE Transactions on Power
  Systems 19~(4) (2004) 1821--1830.

\bibitem{Kim2000}
K.-h. Kim, H.-s. Youn, S.~Member, Y.-c. Kang, {Short-term load forecasting for
  special days in anomalous load conditions using neural networks}, IEEE
  Transactions on Power Systems 15~(2) (2000) 559--565.
\newblock \href {https://doi.org/10.1109/59.867141}
  {\path{doi:10.1109/59.867141}}.

\bibitem{Tiong2008}
J.~Nagi, S.~K. Yap, S.~K. Tiong, S.~K. Ahmed, {Electrical Power Load
  Forecasting using Hybrid Self-Organizing Maps and Support Vector Machines},
  The 2nd International Power Engineering optimization Conference
  (PEOCO)~(June) (2008) 51 -- 56.

\bibitem{Quilumba2014}
F.~L. Quilumba, W.-j. Lee, H.~Huang, D.~Y. Wang, S.~Member, R.~L. Szabados,
  {Using Smart Meter Data to Improve the Accuracy of Intraday Load Forecasting
  Considering Customer Behavior Similarities} (2014) 1--8.

\bibitem{Nguyen2017}
H.~Nguyen, C.~K. Hansen,
  \href{http://ieeexplore.ieee.org/document/7998331/}{{Short-term electricity
  load forecasting with Time Series Analysis}}, 2017 IEEE International
  Conference on Prognostics and Health Management (ICPHM) (2017) 214--221\href
  {https://doi.org/10.1109/ICPHM.2017.7998331}
  {\path{doi:10.1109/ICPHM.2017.7998331}}.
\newline\urlprefix\url{http://ieeexplore.ieee.org/document/7998331/}

\bibitem{Gross1987}
G.~Gross, F.~Galiana, {Short-term load forecasting}, Proceedings of the IEEE
  75~(12) (1987) 1558--1573.
\newblock \href {https://doi.org/10.1109/PROC.1987.13927}
  {\path{doi:10.1109/PROC.1987.13927}}.

\bibitem{Ghofrani}
M.~Ghofrani, M.~Hassanzadeh, M.~S. Fadali, {Smart Meter Based Short-Term Load
  Forecasting for Residential Customers}  13--17.

\bibitem{Fard2014}
A.~K. Fard, M.-R. Akbari-Zadeh, {A hybrid method based on wavelet, ANN and
  ARIMA model for short-term load forecasting}, Journal of Experimental {\&}
  Theoretical Artificial Intelligence 26~(2) (2014) 167--182.
\newblock \href {https://doi.org/10.1080/0952813X.2013.813976}
  {\path{doi:10.1080/0952813X.2013.813976}}.

\bibitem{Humeau2013}
S.~Humeau, T.~K. Wijaya, M.~Vasirani, K.~Aberer, {Electricity load forecasting
  for residential customers: Exploiting aggregation and correlation between
  households}, 2013 Sustainable Internet and ICT for Sustainability, SustainIT
  2013 (2013).
\newblock \href {https://doi.org/10.1109/SustainIT.2013.6685208}
  {\path{doi:10.1109/SustainIT.2013.6685208}}.

\bibitem{Johansson2017}
C.~Johansson, M.~Bergkvist, D.~Geysen, O.~D. Somer, N.~Lavesson, D.~Vanhoudt,
  {Operational Demand Forecasting in District Heating Systems Using Ensembles
  of Online Machine Learning Algorithms}, Energy Procedia 116 (2017) 208--216.

\bibitem{Baram2003}
Y.~Baram, R.~El-Yaniv, K.~Luz, {Online Choice of Active Learning Algorithms},
  Proceedings, Twentieth International Conference on Machine Learning 1 (2003)
  19--26.

\bibitem{Schmitt2008}
J.~Schmitt, M.~Hollick, C.~Roos, R.~Steinmetz, {Adapting the user context in
  realtime: Tailoring online machine learning algorithms to ambient computing},
  Mobile Networks and Applications 13~(6) (2008) 583--598.
\newblock \href {https://doi.org/10.1007/s11036-008-0095-8}
  {\path{doi:10.1007/s11036-008-0095-8}}.

\bibitem{Widmer1996}
G.~Widmer, {Learning in the presence of concept drift and hidden contexts},
  Machine Learning 23~(1) (1996) 69--101.
\newblock \href {https://doi.org/10.1007/bf00116900}
  {\path{doi:10.1007/bf00116900}}.

\bibitem{Pindoriya2008}
N.~M. Pindoriya, S.~N. Singh, S.~K. Singh, {An adaptive wavelet neural
  network-based energy price forecasting in electricity markets}, IEEE
  Transactions on Power Systems 23~(3) (2008) 1423--1432.
\newblock \href {https://doi.org/10.1109/TPWRS.2008.922251}
  {\path{doi:10.1109/TPWRS.2008.922251}}.

\bibitem{GoncalvesDaSilva2014}
P.~{Goncalves Da Silva}, D.~Ilic, S.~Karnouskos, {The Impact of Smart Grid
  Prosumer Grouping on Forecasting Accuracy and Its Benefits for Local
  Electricity Market Trading}, IEEE Transactions on Smart Grid 5~(1) (2014)
  402--410.
\newblock \href {https://doi.org/10.1109/TSG.2013.2278868}
  {\path{doi:10.1109/TSG.2013.2278868}}.

\bibitem{Gzik2014}
K.~Crammer, O.~Dekel, J.~Keshet, S.~Shalev-Shwartz, Y.~Singer, {Online
  Passive-Aggressive Algorithms}, Journal of Machine Learning Research (2006).
\newblock \href {https://doi.org/10.1201/b15810-63}
  {\path{doi:10.1201/b15810-63}}.

\bibitem{Box1964}
G.~E.~P. Box, D.~Cox, {An Analysis of Transformations}, Journal, Source
  Statistical, Royal Series, Society 26~(2) (1964) 211--252.

\bibitem{forgy65}
E.~Forgy, Cluster analysis of multivariate data: Efficiency versus
  interpretability of classification, Biometrics 21~(3) (1965) 768--769.

\bibitem{Hinton1989}
G.~E. Hinton, {Connectionist learning procedures}, Artificial Intelligence
  40~(1-3) (1989) 185--234.
\newblock \href {https://doi.org/10.1016/0004-3702(89)90049-0}
  {\path{doi:10.1016/0004-3702(89)90049-0}}.

\bibitem{Tibshirani1996a}
R.~Tibshirani, {Regression Shrinkage and Selection Via the Lasso}, Journal of
  the Royal Statistical Society: Series B (Methodological) 58~(1) (1996)
  267--288.
\newblock \href {https://doi.org/10.1111/j.2517-6161.1996.tb02080.x}
  {\path{doi:10.1111/j.2517-6161.1996.tb02080.x}}.

\bibitem{GeladiPaul1994Mrac}
P.~Geladi, Measurement, regression and calibration, philip brown, oxford
  statistical science series. vol. 12, oxford science publications, oxford,
  1993, no of pages: 224. price: £27.50. isbn 0 19-852245-2, Journal of
  chemometrics 8~(5) (1994) 371--372.

\bibitem{Geostatistics2010}
J.~Friedman, T.~Hastie, R.~Tibshirani, {Regularization Paths for Generalized
  Linear Models via Coordinate Descent}, Journal of Statistical Software 33~(1)
  (2010) 1--22.
\newblock \href {http://arxiv.org/abs/0908.3817} {\path{arXiv:0908.3817}},
  \href {https://doi.org/10.1016/j.expneurol.2008.01.011}
  {\path{doi:10.1016/j.expneurol.2008.01.011}}.

\bibitem{Fike1988}
B.~Efron, T.~Hastie, I.~Johnstone, R.~Tibshirani, {Least angle regression}, The
  Annals of Statistics 32~(2) (1988) 440--444.
\newblock \href {https://doi.org/10.1109/glocom.1988.25879}
  {\path{doi:10.1109/glocom.1988.25879}}.

\bibitem{Breiman2001}
L.~Breiman, {Random forests}, Machine Learning 45~(1) (2001) 5--32.
\newblock \href
  {http://arxiv.org/abs//dx.doi.org/10.1023{\%}2FA{\%}3A1010933404324}
  {\path{arXiv:/dx.doi.org/10.1023{\%}2FA{\%}3A1010933404324}}, \href
  {https://doi.org/10.1023/A:1010933404324}
  {\path{doi:10.1023/A:1010933404324}}.

\bibitem{Freund1997}
Y.~Freund, R.~E. Schapire, {A Decision-Theoretic Generalization of On-Line
  Learning and an Application to Boosting}, Journal of Computer and System
  Sciences 55~(1) (1997) 119--139.
\newblock \href {https://doi.org/10.1006/jcss.1997.1504}
  {\path{doi:10.1006/jcss.1997.1504}}.

\bibitem{316}
J.~H. Friedman, {Greedy Function Approximation: A Gradient Boosting Machine}
  (316) 400.

\bibitem{Cortes1995}
C.~Cortes, V.~Vapnik, {Support-Vector Networks}, Machine Learning 20~(3) (1995)
  273--297.
\newblock \href {http://arxiv.org/abs/arXiv:1011.1669v3}
  {\path{arXiv:arXiv:1011.1669v3}}, \href
  {https://doi.org/10.1023/A:1022627411411}
  {\path{doi:10.1023/A:1022627411411}}.

\bibitem{scikit-learn}
F.~Pedregosa, G.~Varoquaux, A.~Gramfort, V.~Michel, B.~Thirion, O.~Grisel,
  M.~Blondel, P.~Prettenhofer, R.~Weiss, V.~Dubourg, J.~Vanderplas, A.~Passos,
  D.~Cournapeau, M.~Brucher, M.~Perrot, E.~Duchesnay, Scikit-learn: Machine
  learning in {P}ython, Journal of Machine Learning Research 12 (2011)
  2825--2830.

\bibitem{creme}
{Creme}, https://pypi.org/project/creme/ (2019).

\bibitem{Quinlan}
J.~R. Quinlan, C4.5: Programs for Machine Learning, Morgan Kaufmann Publishers
  Inc., San Francisco, CA, USA, 1993.

\bibitem{Shu2006}
F.~Shu, C.~Luonan, {Short-term load forecasting based on an adaptive hybrid
  method}, Power Systems, IEEE Transactions on 21~(1) (2006) 392--401.
\newblock \href {https://doi.org/10.1109/TPWRS.2005.860944}
  {\path{doi:10.1109/TPWRS.2005.860944}}.

\bibitem{Drucker1997}
H.~Drucker, C.~J.~C. Burges, L.~Kaufman, A.~Smola, V.~Vapnik, {Support vector
  regression machines}, Advances in Neural Information Processing Systems 1
  (1997) 155--161.
\newblock \href {https://doi.org/10.1.1.10.4845} {\path{doi:10.1.1.10.4845}}.

\bibitem{Smola2004}
A.~J. Smola, B.~Sch{\"{o}}lkopf, {A tutorial on support vector regression},
  Statistics and Computing 14~(3) (2004) 199--222.
\newblock \href {http://arxiv.org/abs/arXiv:1011.1669v3}
  {\path{arXiv:arXiv:1011.1669v3}}, \href
  {https://doi.org/10.1023/B:STCO.0000035301.49549.88}
  {\path{doi:10.1023/B:STCO.0000035301.49549.88}}.

\bibitem{Akaike1974}
H.~Akaike, {A New Look at the Statistical Model Identification}, IEEE
  Transactions on Automatic Control 19~(6) (1974) 716--723.
\newblock \href {http://arxiv.org/abs/arXiv:1011.1669v3}
  {\path{arXiv:arXiv:1011.1669v3}}, \href
  {https://doi.org/10.1109/TAC.1974.1100705}
  {\path{doi:10.1109/TAC.1974.1100705}}.

\bibitem{Pao2007}
H.-T. Pao, {Forecasting electricity market pricing using artificial neural
  networks}, Energy Conversion and Management 48~(3) (2007) 907--912.
\newblock \href {https://doi.org/10.1016/j.enconman.2006.08.016}
  {\path{doi:10.1016/j.enconman.2006.08.016}}.

\bibitem{Ma2009}
J.~Ma, L.~K. Saul, S.~Savage, G.~M. Voelker, {Identifying suspicious URLs: An
  application of large-scale online learning}, Proceedings of the 26th
  International Conference On Machine Learning, ICML 2009 (2009) 681--688.

\bibitem{dukes_511}
{Department for Business and Industrial Strategy, UK Government}, Power
  stations in the united kingdom, may 2019, Digest of United Kingdom Energy
  Statistics (DUKES) (2019).

\bibitem{companies_house}
{Department for Business, Energy \& Industrial Strategy}, Companies house -
  gov.uk, UK Government (2019).

\bibitem{DBEIS2019}
{Department for Business Energy {\&} Industrial Strategy}, {Updated energy and
  emissions projections 2018}, The Energy White Paper~(April) (2019).

\bibitem{gbnationalgridstatus_2019}
{Elexon portal and Sheffield University}, G.b. national grid status,
  https://www.gridwatch.templar.co.uk/.

\bibitem{ESO2019}
{National Grid}, {STOR Market Information Report}~(October) (2019) 0--11.

\end{thebibliography}


%
%
%
\end{document}